\documentclass[showpacs,preprintnumbers,english,amsmath,twocolumn,amssymb]{revtex4}
\usepackage[T2A,T1]{fontenc}
\usepackage[utf8]{inputenc}
\usepackage{graphicx}
\usepackage{amssymb}
\usepackage{verbatim}
\usepackage{epic,eepic}
\usepackage{babel}
\usepackage{color}
\usepackage{diagbox}

\usepackage[normalem]{ulem}  

\renewcommand\sout{\bgroup \color{red} \ULdepth=-.5ex \ULset}

\begin{document}

\title{Universal Energy Dependence of Measured Temperatures for Baryons Produced in  Heavy-ion Collisions}

\author{Lilin Zhu$^{1}$, Hua Zheng$^{2}$\footnote{Corresponding author: zhengh@snnu.edu.cn}, Ke Da$^{2}$, Huanjing Gong$^{1}$, Zhizhen Ye$^{1}$, Guiqi Liu$^{1}$, Rudolph C. Hwa$^{3}$}
\affiliation{
$^{1}$Department of Physics, Sichuan University, Chengdu 610064, China;\\
$^{2}$School of Physics and Information Technology, Shaanxi Normal University, Xi'an 710119, China;\\
$^{3}$Institute of Fundamental Science, University of Oregon, Eugene, Oregon 97403-5203, USA.
}

\begin{abstract}
From the data on baryon production in heavy-ion collisions it is shown that a set of measurable functions exist, one for each baryon type, that depend exponentially on transverse momenta up to the maximum values detected for all collision energies ranging from the low end of RHIC BES to the high end of CERN LHC. The implied temperatures satisfy a scaling law in collision energy with a universal exponent for all baryon types which is the novel discovery. A self-similar thermal source is implied. Furthermore, it is shown how the scaling behavior depends on centralities. Those features in the data are related to simple properties of light and strange quarks by use of the recombination model. Prediction of $\phi$ meson production is made and then verified by existing data. 
\end{abstract}





\maketitle

\section{introduction}
There exist now data on baryon production in heavy-ion collisions for a wide range of collision energy from $\sqrt{s_{NN}} = 7.7$ GeV to 5.02 TeV \cite{STAR:2017sal, STAR:2019bjj, Adam:2015kca, Abelev:2014uua, Abelev:2013xaa, ABELEV:2013zaa, Acharya:2019yoi, PHENIX:2013kod, STAR:2006egk, STAR:2007zea, STAR:2010yyv, STAR:2008bgi}. We show that they can be linked by a scaling behavior in the entire range. That scaling law is rooted in the data  without modeling, and is characterized by a universal scaling exponent, a pure number. No part of our analysis of the data relies on any theoretical models. The main purpose of this paper is to exhibit those phenomenological facts in the data that has hitherto been unrecognized. It seems that the  behavior cannot easily be explained in the prevailing theoretical understanding of the physics of heavy-ion collisions. We can offer no satisfactory explanation,  but regard the new discovery worthy of presentation on its own merits, as it deserves a collective effort to evaluate its significance and implications.
 
In Ref. \cite{Hwa:2018qss} we have reported on certain features of the data on baryon production in heavy-ion collisions that showed some regularity at energies $\sqrt{s_{NN}} =$ 0.0624, 0.2 and 2.76 TeV  \cite{Adam:2015kca,  Abelev:2013xaa, ABELEV:2013zaa, PHENIX:2013kod, STAR:2006egk, STAR:2007zea, STAR:2010yyv}. Some part of the measurement has been extended to 5.02 TeV \cite{Acharya:2019yoi}. Now, with BES data available \cite{STAR:2017sal, STAR:2019bjj, PHENIX:2013kod, STAR:2006egk, STAR:2007zea, STAR:2010yyv} we find that the same features persist down to 7.7 GeV, as we shall describe below. The key to showing such features is in the use of a function constructed out of the measured transverse-momentum ($p_T$) distribution. It is defined as
\begin{equation}
B_h(s, p_T)=\frac{m_T^h}{p_T^2} \left[\frac{dN_h}{2\pi p_T dp_T dy}(s, p_T)\right].  \label{1}
\end{equation}
The quantity inside the square brackets is the experimentally measured  inclusive distribution of hadron $h$ at all azimuthal angle $\varphi$ and for rapidity $y$ in the mid-range.
We use $h$ to denote baryons ($b$) or anti-baryons ($\bar b$)  with $b$ standing for ($p, \Lambda, \Xi, \Omega$) collectively, or any one of the four, specifically. We shall use $\sqrt{s}$ to denote $\sqrt{s_{NN}}$ in units of TeV, i.e.,  
\begin{equation}
\sqrt{s} = \sqrt{s_{NN}}/1{\rm TeV} ,  \label{2}
\end{equation}
so that it is a dimensionless variable that is more suitable in an expression of power-law behavior in $\sqrt{s}$. The transverse mass is $m_T^h = (m_h^2 + p_T^2)^{1/2}$, $m_h$ being the mass of $h$. 
The pre-factor in front of the square brackets in Eq.(\ref{1}) has root elsewhere \cite{Hwa:2018qss}, for which we do not digress here to explain. For the present it should be understood that the pre-factor does not alter the character that $B_h(s, p_T)$ is an experimentally determinable quantity. The measured distributions, of course, depend on centrality. 
We consider first the most central bin ($0-5\%$ or wider, as indicated), and then take up  the subject of centrality dependence after the scaling law is established.  

Furthermore, these properties can emerge naturally in the recombination model (RM) connecting to the simple properties of light and strange quarks. It should be pointed out that the recombination model that we used to enlighten the hadronization part of the problem is not essential to our presentation of the universal behavior in the data, but useful in relating it to parton distributions. We also take $\phi$ meson to demonstrate that the study presented in this paper is on the right track $-$ by a prediction that can be actually verified by existing data.

The paper is organized as follows. In Sec. \ref{sec2}, we present the empirical properties of the baryons and anti-baryons spectra for various centralities at BES energies as well as LHC and establish the scaling law of the measured temperatures for them. We show how those properties emerge naturally in the framework of recombination model and the results of baryon spectra from RM in Sec. \ref{sec3}. In Sec. \ref{sec4}, we show some discussions on the results and the prediction of $\phi$ production. Concluding remarks are made in the final section.
	
\section{notable features of (anti)baryon spectra}
\label{sec2}
\begin{figure}[hpt]
\centering
	\includegraphics[width=0.4\textwidth]{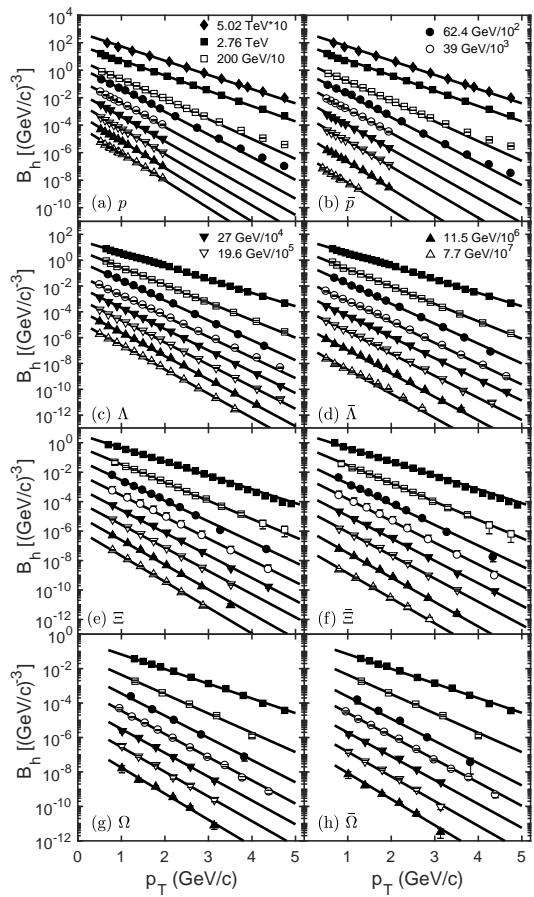}
	\caption{(Left panels) Baryon spectra function $B_h(p_T, s)$; (Right panels) Anti-baryons spectra function  $B_{\bar h}(p_T, s)$. The centralities are 0-5\% for all collision energies except the following: 0-10\%  for $p$ and $\bar{p}$ at $\sqrt{s_{NN}}=62.4$ and 200 GeV,  for $\Xi$ and $\bar{\Xi}$ at 2.76 TeV,  and for $\Omega$ and $\bar{\Omega}$ at 11.5, 19.6, 27, 39 GeV and 2.76 TeV,  and 0-20\%  for $\Omega$ and $\bar{\Omega}$ at 62.4 GeV. The lines are the best fits with Eq.(\ref{3}) for the region $p_T<3$ GeV/c.}
	\label{fig1}
\end{figure}

Using the data from Refs. \cite{Acharya:2019yoi, Adam:2015kca, Abelev:2014uua, Abelev:2013xaa, ABELEV:2013zaa,  STAR:2017sal, STAR:2019bjj, PHENIX:2013kod, STAR:2006egk, STAR:2007zea, STAR:2010yyv, STAR:2008bgi} for the distributions in the square brackets in Eq. (\ref{1}), we obtain the points in Fig. \ref{fig1} for the most central bins ($0-5\%$ or wider, as indicated) at all energies $\sqrt{s_{NN}}$ where data exist. Each of the eight subfigures displays $B_h(s, p_T)$ for $h$ being $b$ or $\bar b$. The error bars, if not visible, are smaller than the sizes of the symbols. The data at 2.76 and 5.02 TeV are for $p+\bar p$ \cite{ Acharya:2019yoi, Adam:2015kca}. Since our focus in the paper will be on the $p_T$ dependence of $B_h(s, p_T)$, not on its magnitude, we assume here that the data for $p$ and $\bar p$ separately are half of the published data for $p+\bar p$ so that they can appear in Fig. \ref{fig1} separately. The same is true with how we treat the data at 200 GeV for $\Omega+\bar\Omega$ \cite{STAR:2006egk}: they are split equally for presentation in Fig. \ref{fig1}. The centrality bins are indicated in the figure caption. 
The straight lines are fits to be discussed below. It is remarkable how all points line up so well  along straight lines for all hadron types and for almost all energies. Slight deviations from straight lines at high $p_T$ for $p$ and $\bar p$ at 62.4 and 200 GeV are presumably due to the effects of hard parton scattering, but why they do not show up for higher energies is puzzling. Our concern in this paper is on the universal properties at lower $p_T$, so we leave aside those issues  here.

We fit the points in the region $p_T<3$ GeV/c by the solid lines shown. Those lines are extended to higher $p_T$ to show how well all the hyperon data fall on them. Quantitative analyses below are based only on the slopes of those lines determined by the $p_T<3$ GeV/c region. Specifically, we fit the distributions by the equation
\begin{equation}
B_h(s, p_T)=A_h(s)\exp [-p_T/T_h(s)] .  \label{3}
\end{equation}
We refer to the inverse slopes $T_h(s)$ as {\it measured} temperatures. In using Eq.(\ref{3}) to fit the data in Fig.\ 1, we have not relied on any model; thus $T_h(s)$ are measured quantities. Their values are given in Table \ref{tab1}.

\begin{table}
\centering
\renewcommand{\arraystretch}{1.5}
\tabcolsep0.02in 
\begin{tabular}{c|cccccccc}
\hline
\hline
\diagbox[width=6em, height=4em,innerrightsep=0.4cm, innerleftsep=-0.05cm]{$\sqrt{s_{NN}}$}{$T_h$}{$h$} &$p$ &$\bar p$ &$\Lambda$ &$\bar\Lambda$ &$\Xi$ &$\bar\Xi$ &$\Omega$ &$\bar\Omega$\\ 
 \hline
5020 & 0.42 & 0.42&\\
2760 &0.39 &0.39 &0.423 &0.423 &0.463 &0.463 &0.51 & 0.51\\
200   &0.3 &0.302 &0.32 &0.322 &0.351 &0.351 &0.387 &0.387\\
62.4  &0.263 &0.264 &0.282 &0.282 &0.311 &0.31 &0.337 &0.337\\
39 &0.25 &0.25 &0.272 &0.27 &0.295 &0.293 & 0.322 &0.322 \\
27 &0.243& 0.243 &0.263 &0.265 & 0.283 &0.285 &0.314 &0.314 \\
19.6 &0.233 &0.234 &0.254 &0.254 &0.276 &0.276 &0.303 &0.303 \\
11.5 &0.222 &0.222 &0.24 &0.24 &0.26 &0.261 &0.287 &0.287\\
7.7 &0.217 &0.217 &0.232  &0.232 &0.247 &0.255 &0.28 &0.28 \\
 \hline
  \hline
 \end{tabular}
 \caption{Values of $T_h(s)$ in GeV/c determined from Fig. \ref{fig1} for listed $\sqrt {s_{NN}}$ in GeV vertically and $h$-type horizontally.}
 \label{tab1}
 \end{table}
 
 \begin{figure}[hpt]
\centering
	\includegraphics[width=0.35\textwidth]{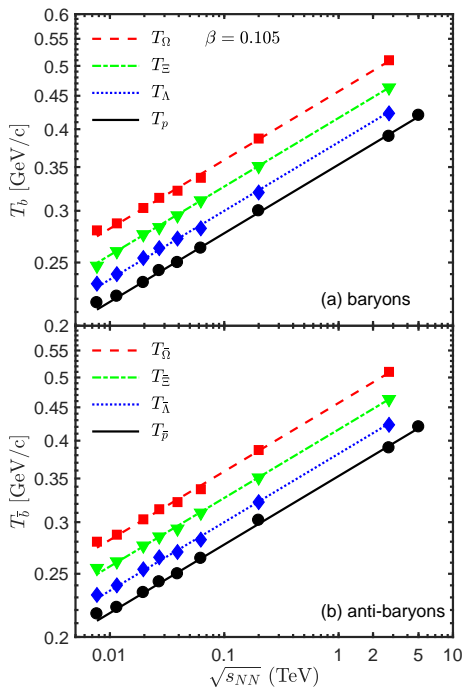}
	\caption{Measured temperatures $T_b(s)$ in (a) and $T_{\bar b}(s)$ in (b) at different collision energies. The lines are from a simultaneous fit of all points using Eq.(\ref{4}) with a common $\beta$, excluding the $p$ and $\bar p$ points at $\sqrt{s_{NN}}=7.7$ GeV.}
	\label{fig2}
\end{figure}

 \begin{figure*}[t]
\centering
	\includegraphics[width=0.7\textwidth]{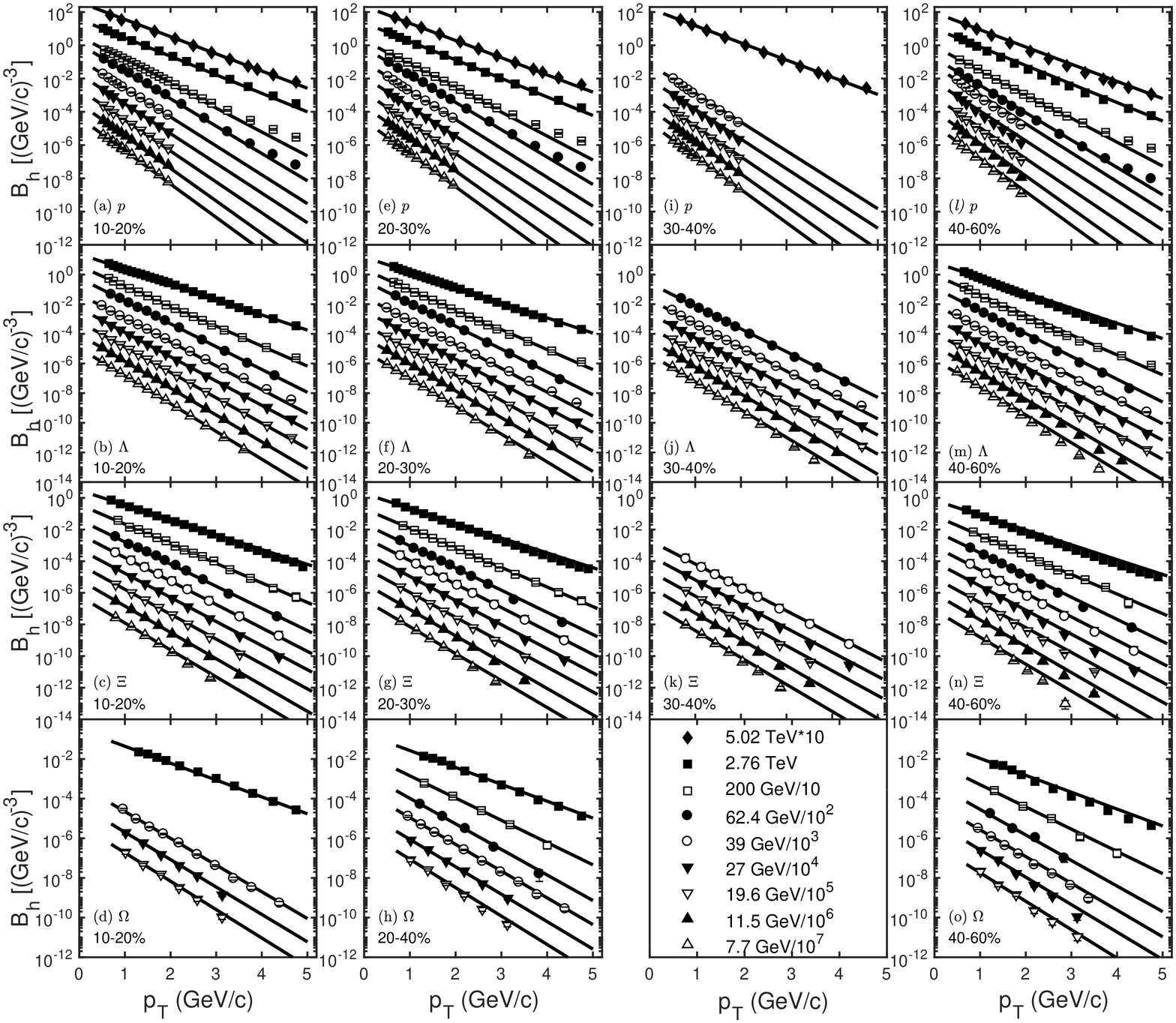}
	\caption{Baryon spectra function $B_h(s, p_T)$ at different collision centralities and collision energies as indicated, except for the following: in (e) 20-40\% at 62.4, 200 GeV and 2.76 TeV; in (f) 20-40\% at 200 GeV and 2.76 TeV; in (g) 20-40\% at 62.4, 200 GeV and 2.76 TeV; in ({\it l}) 40-80\% at 62.4 GeV. The normalizations of straight lines are adjusted to fit the lowest two $p_T$ points, but their slopes are as specified in Table \ref{tab1}.}
	\label{fig3}
\end{figure*}

\begin{figure*}[t]
\centering
	\includegraphics[width=0.7\textwidth]{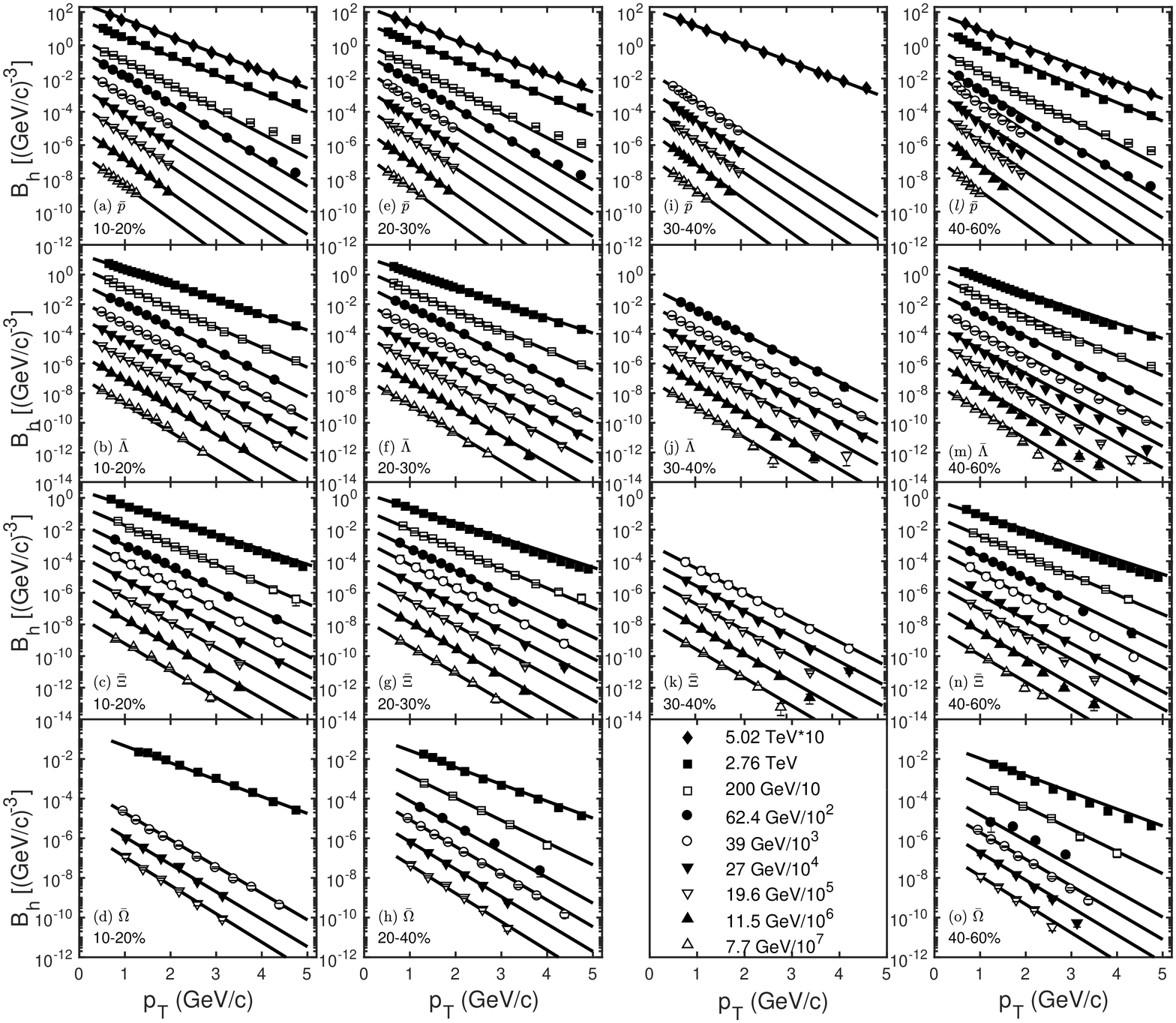}
	\caption{Same as for Fig. \ref{fig3} but for anti-baryons.}
	\label{fig4}
\end{figure*}

In Fig. \ref{fig2} we show the values in Table \ref{tab1} in a log-log plot of $T_h(s)$ vs $\sqrt{s_{NN}}$. We fit all the points of each $h$ type by a straight line, except the two lowest points of $p$ and $\bar p$. In the fit we demand that all the lines have the same slope, with a result that justifies that demand, since all the lines go through all the points. The eight lines are summarized by one equation
\begin{equation}
T_h(s) = T_h(1) \sqrt{s}^\beta , \qquad \beta=0.105 ,   \label{4}
\end{equation}
in terms of the dimensionless variable $\sqrt{s}$ defined in Eq. (\ref 2). The parameters $T_h(1)$ are adjusted for each $h$, while $\beta$ is the universal scaling exponent determined by the best fit for all $h$. The power-law behavior in Eq.(\ref 4) is notable in that it is valid over  a wide range of  three orders of magnitude in $\sqrt{s}$ for all $h$, and that $\beta$ is independent of $h$.  Again, the result is contained entirely in the data without any model input. This is an unexpected phenomenological property never seen before in any other areas of high- or low-energy collisions. 
In particular, it is known not to exist in $pp$ and $p\bar p$ collisions. Thus, there must  be dependence on centrality, a topic to be discussed below.

\begin{figure}[hpt]
\centering
	\includegraphics[width=0.4\textwidth]{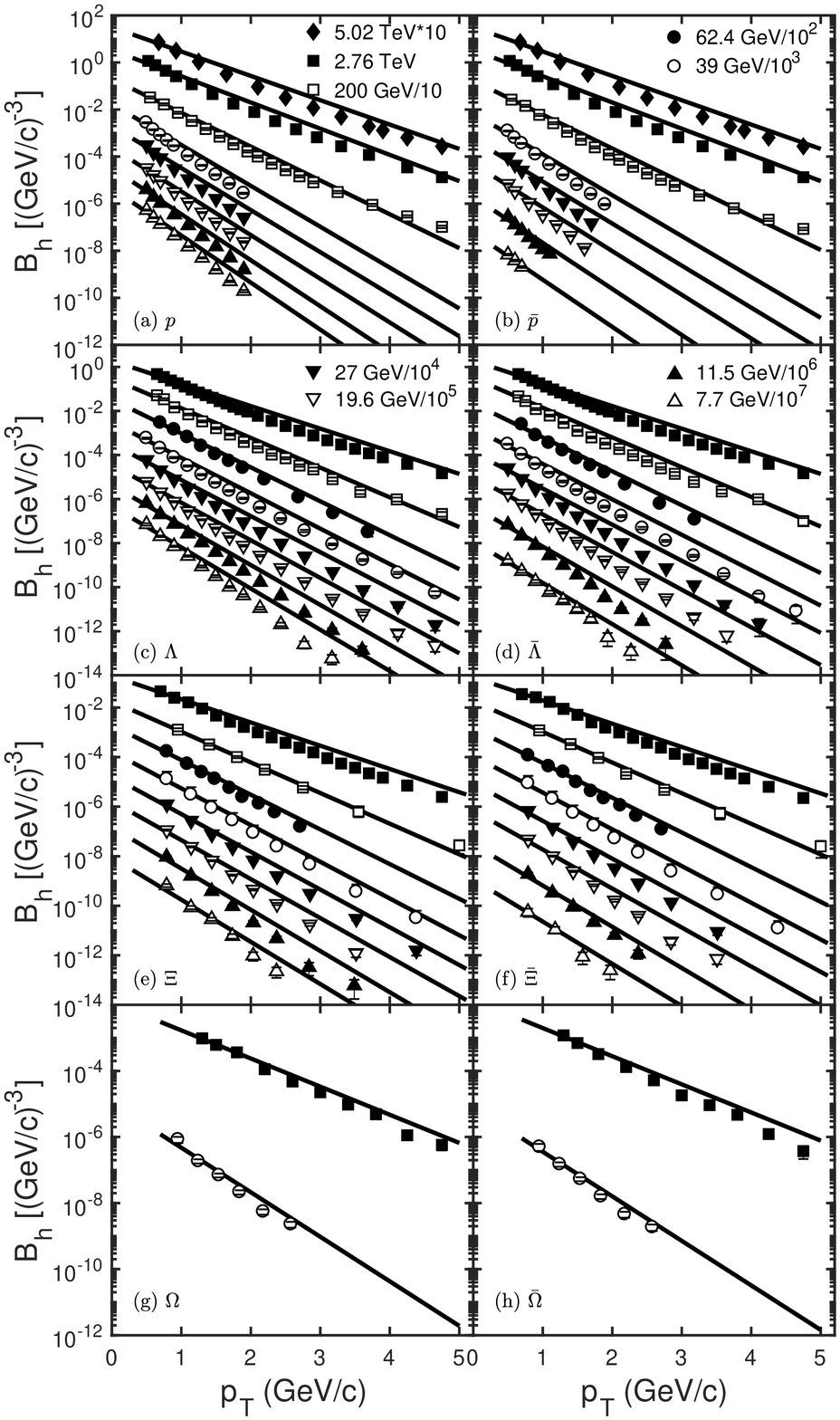}
	\caption{Baryon spectra function $B_h(s, p_T)$ at different collision energies for the centrality of 60-80\%, except 60-92\% for $p$ and $\bar p$ at $\sqrt{s_{NN}}=200$ GeV. The normalizations of straight lines are adjusted to fit the lowest two $p_T$ points, but their slopes are as specified in Table \ref{tab1}.}
	\label{fig5}
\end{figure}

\begin{figure*}[pht]
 \includegraphics[width=0.7\textwidth]{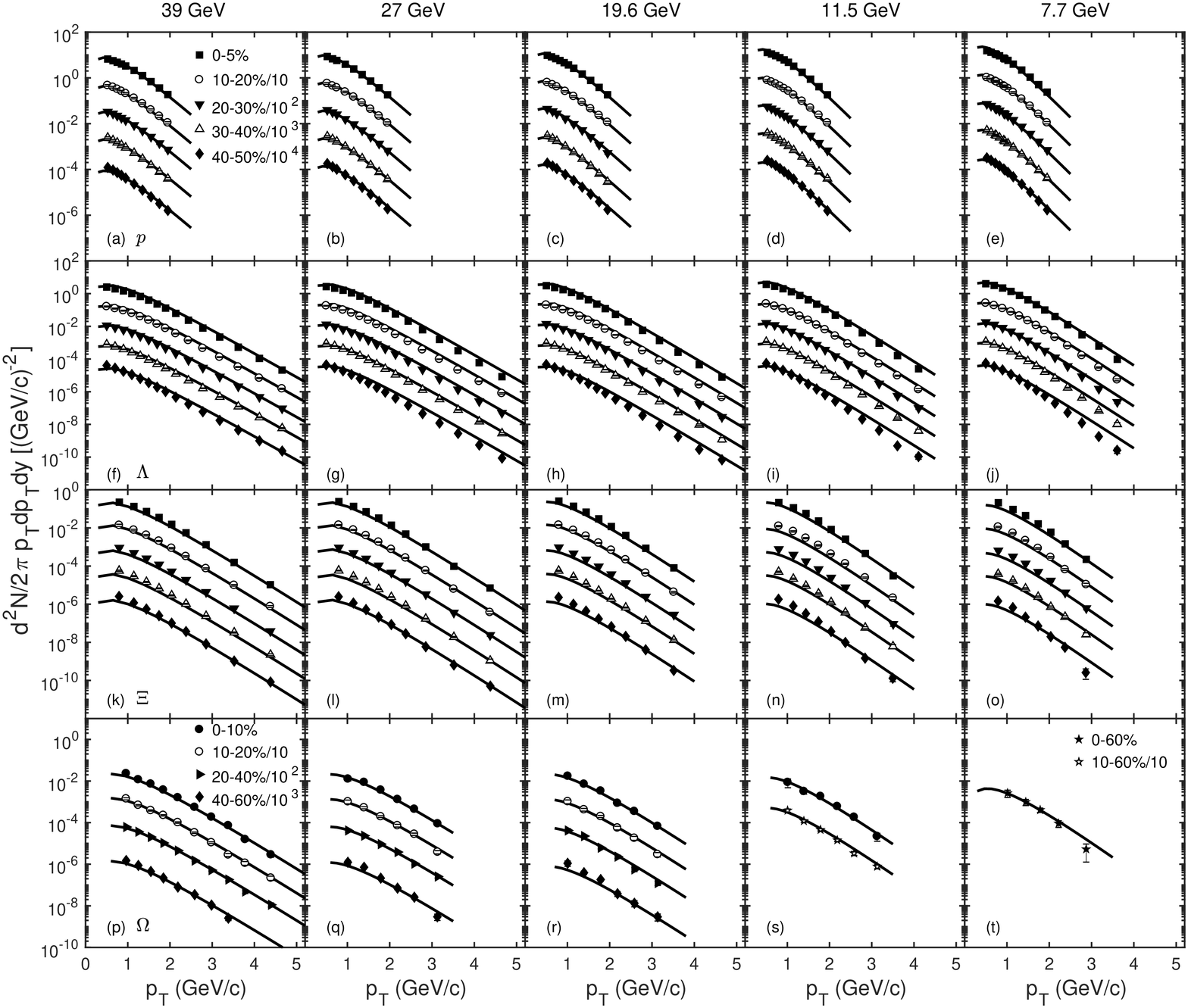}
 \caption{Transverse momentum spectra of baryons ($p$, $\Lambda$, $\Xi$, $\Omega$) from the recombination model in Au+Au collisions at $\sqrt{s_{NN}}=39$, 27, 19.6, 11.5 and 7.7 GeV for different centrality classes as indicated, except 40-60\% for $\Lambda$ and $\Xi$ (full diamonds) in the subfigures from (f) to (o). The spectra for centralities other than 0-5\% (0-10\%) are scaled for clarity as shown in the figure. The experimental data are taken from Refs. \cite{STAR:2017sal, STAR:2019bjj}.}
\label{fig6}
\end{figure*}

The values of $T_h(1)$ will not be displayed here, except for $h=p$ and $\Omega$ that will be needed below. They are notationally abbreviated by $T_{1,2}$
\begin{eqnarray}
T_1=T_p(1)=0.353 {\rm GeV/c},  \quad		
T_2=T_\Omega(1)=0.457 {\rm GeV/c} . \label{5a}
\end{eqnarray}

Scaling behavior such as the power law in Eq. (\ref{4}) is generally referred to as self-similarity. 
In geometrical patterns they are called fractals. 
Here, it suggests that the dynamical origin of the production of those heavy hadrons is invariant under the change of collision energy. That seems to depart from our conventional understanding of the formation of quark gluon plasma in heavy-ion collisions. The observables usually studied are on pions and other low-mass mesons, which are abundantly produced and display the properties of a hot and dense medium whose expansion can be investigated in hydrodynamics \cite{Schnedermann:1993ws, Cassing:1999es, Song:2017wtw, Gale:2013da, Zhao:2017yhj, Li:2022pyw}. 
There is no invariance in $\sqrt{s}$ that is generally known. Our results seem to suggest that the production of baryons and anti-baryons originates from the core of the hot medium that has universal features and differs from the part where pions are produced. Exponential behavior in $p_T$ is traditionally regarded as evidence of thermal source. Here we have found indications that the thermal system possesses properties of self-similarity, although we have not yet established a connection between the thermal source and the observed hadrons. That connection is the problem of hadronization.
 
We now consider the centrality dependence of the universal behavior expressed by Eqs. (\ref{3}, \ref{4}). The data points are shown in Fig. \ref{fig3} for baryons and Fig. \ref{fig4} for anti-baryons. Horizontally, the subfigures show centrality varying from 10-20\% to 40-60\%. Vertically, the subfigures show hadron type changing from $p$ to $\Omega$ and from $\bar p$ to $\bar \Omega$. In each subfigure the collision energies vary over the entire range, as indicated in the subfigure for symbols, but for some hadrons the data are not yet available for certain centralities and energies. The straight lines are plots of  Eq. (\ref{3}) with $T_h(s)$ as given in Table \ref{tab1} but with their normalizations adjusted to fit the lowest two $p_T$ points of each set. We observe that while the lines remain to be excellent fits for centrality bins 10-20\% and 20-30\%, they begin to miss some data points at higher $p_T$ for more non-central bins, especially with hyperons at lower energies. For 40-60\% all data points have steeper slopes at RHIC energies and below, compared to the straight lines. In Fig. \ref{fig5} we show the very peripheral collisions at 60-80\% where all data points at all energies have steeper trend than the lines. Clearly, the implication is that the universality behavior deteriorates in non-central collision beginning at 30\%. Steeper slope means lower temperature. That is a reasonable property for smaller systems generated by less overlap of the colliding nuclei.

\section{quark recombination}
\label{sec3}
The thermal source that is relevant to our observables cannot be a hadron gas, since protons and $\Omega$s do not interact effectively to maintain thermal contact. 
The fact that the values of the measured temperatures, $T_h$, depend so regularly on the strangeness of the baryon type suggests that the thermal source must be at the partonic level, where only two types of quarks (light and strange) need to be considered here and they can interact effectively through gluon exchange, pair annihilation and creation. The hadronization of quarks to form baryons at intermediate $p_T$ can be well described by the recombination/coalescence models \cite{Das:1977cp, rm1, rm2, Molnar:2003ff, Fries:2003vb, Greco:2003xt}, which have been successful in explaining the large $p/\pi$ ratio \cite{STAR:2007zea, rm1, STAR:2011iap, PHENIX:2003wtu,  STAR:2006uve, ALICE:2014juv}.

The invariant distribution in the recombination model (RM) \cite{Das:1977cp, rm1, rm2}, averaged over all $\varphi$ at mid-rapidity, is
\begin{equation}
p^0{d{\bar N}_h\over dp_T} = \int \left(\prod_{i=1}^3 {dp_i\over p_i}\right)F(p_1,p_2,p_3)R_h(p_1,p_2,p_3,p_T) ,	\label{5}
\end{equation}
where $F(p_1, p_2, p_3)$ is the parton distribution of the three quarks (with momenta $p_i$) that are to recombine through the function $R_h(p_1,p_2,p_3,p_T)$ in producing the hadron $h$. For a thermal source of the partons, we assume a factorizable distribution
\begin{equation}
F(p_1,p_2,p_3) = \prod_{j\in h} p_i{dN_j\over dp_i} = \prod_{j\in h} C_jp_i\exp{(-p_i/T_j)} , \label{6}
\end{equation}
where the products are for $j$ to range over three terms depending on the $h$-type, i.e., $h=\{qqq,qqs,qss,sss\}$ for the strangeness number of $h$ being $n_s=\{0,1,2,3\}$, respectively; each of the three terms depends successively on $p_i$ with $i=\{1,2,3\}$. $T_j$ can be either $T_q$ or $T_s$. For the recombination function we adopt the simplest form on the assumption that each of the quarks in $h$ has momentum 1/3 of the hadron
\begin{equation}
R_h(p_1,p_2,p_3, p_T) = g_h \prod_{i=1}^3 \delta\left(\frac{p_i}{p_T}-{1\over 3}\right),   \label{7}
\end{equation}
with $g_h$ being just a constant. Substituting Eqs. (\ref{6}) and (\ref{7}) into (\ref{5}) we obtain
\begin{equation}
p^0{d{\bar N}_h\over dp_T} = A_h p_T^3 e^{-p_T/T_h},	\label{8}
\end{equation}
where $A_h=g_h\Pi_jC_j$ and 
\begin{equation}
{1\over T_h} = {1\over 3} \sum_j {1\over T_j} = {1\over 3} \left({3-n_s\over T_q} + {n_s\over T_s}\right) .	\label{9}
\end{equation}
For $|y|<0.5$ we replace $p^0$ by $m_T^h$, and get 
\begin{equation}
\frac{d{\bar N}_h}{p_T dp_T} = \frac{p_T^2}{m_T^h} A_h e^{-p_T/T_h} . 	\label{10}
\end{equation}

\begin{table}
\centering
\renewcommand{\arraystretch}{1.5}
\tabcolsep0.03in 
\begin{tabular}{cccc}
\hline
 \hline
 $\sqrt{s_{NN}}$ [GeV] & centrality  &  $C$ [(GeV/c)$^{-1}$]& $C_s$  [(GeV/c)$^{-1}$]\\ 
 \hline
  &0-5\%  &38.56&11.63 \\
&10-20\% &31.88 &10.26 \\
39&20-30\%  &28.85 &7.51 \\
&30-40\%  &24.74 &6.14 \\
&40-60\%  &18.68 &4.20 \\
 \hline
  &0-5\%  &43.44 &11.63 \\
&10-20\% &36.32 &9.97 \\
27&20-30\%  &30.64 &8.04 \\
&30-40\%  &25.97 &5.94 \\
&40-60\%  & 22.11 &5.04 \\
 \hline
  &0-5\%  &45.71 &10.50 \\
&10-20\% &39.12 &8.81\\
19.6&20-30\%  &33.71 &7.36 \\
&30-40\%  &27.32 &6.23\\
&40-60\%  &23.77 &4.48 \\
 \hline
  &0-5\%  &55.08 &10.93 \\
&10-20\% &45.73&7.66 \\
11.5&20-30\%  &39.10 &6.42 \\
&30-40\%  &32.99&5.42\\
&40-60\%  &26.76 &3.43 \\
&10-60\%  &     &7.66 \\
 \hline
  &0-5\%  &60.98 &9.07\\
&10-20\% &51.15 &7.51 \\
7.7&20-30\%  &42.45 &5.94\\
&30-40\%  &37.45&4.97\\
&40-60\%  &29.74 &3.25 \\
&0-60\%    &       &7.40\\
 \hline
  \hline
 \end{tabular}
 \caption{Parameters $C$ and $C_s$ for Au+Au collisions at $\sqrt{s_{NN}}=39$, 27, 19.6, 11.5 and 7.7 GeV, respectively.} 
 \label{tab2}
 \end{table}

This is to be identified with the experimental quantity inside the square brackets in Eq. (\ref{1}). We now see the  origin of the prefactor in Eq. (\ref{1}) that results in Eq. (\ref{3}). Note that in the use of RM above we have assumed validity at any $s$ so long as the parton model is valid. The hadron production from Au+Au at $\sqrt{s_{NN}}=62.4$, 200 GeV and Pb+Pb at  $\sqrt{s_{NN}}=$2.76 and 5.02 TeV have been studied in our earlier works \cite{rm1, rm2, Zhu:2021fbs}. Here, we extend the investigation to the lower collision energies. For baryon production the component thermal-thermal-thermal (TTT) recombination is prevalent at $p_T<2$ GeV/c which corresponds to Eq. (\ref{10}). Figure \ref{fig6} shows our results for the transverse momentum spectra of four baryons, i.e., $p$, $\Lambda$, $\Xi$ and $\Omega$, in Au+Au collisions at $\sqrt{s_{NN}}=39$, 27, 19.6, 11.5 and 7.7 GeV for various centrality classes, respectively. We emphasize that for thermal partons the inverse slopes $T_q$ and $T_s$ are independent of centrality. So the unknown parameters are just the two normalization factors, $C$ and $C_s$, whose values are given in Table. \ref{tab2}, to describe the four baryons simultaneously for each centrality.  Evidently, the agreement with the data for $p$, $\Lambda$, $\Xi$ and $\Omega$ in Fig. \ref{fig6} is good within 2 GeV/c. For higher $p_T$, the other recombination components should be considered. Again, the success of the RM implies that the thermal source relevant to our observables is at the partonic level.

\section{discussion and a prediction}
\label{sec4}
\begin{figure}[hpt]
\centering
	\includegraphics[width=0.48\textwidth]{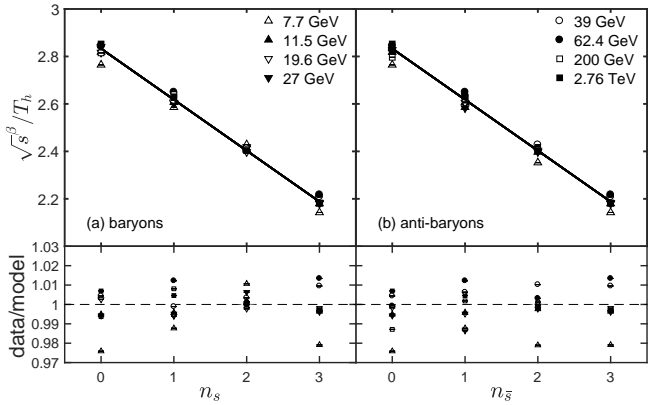}
	\caption{$\sqrt{s}^\beta/T_h(s)$ vs (a) number of $s$ quarks in $b$ and (b) of $\bar s$ anti-quarks in $\bar b$. The data points are from Table I, while the line is drawn according to Eq.(\ref{11}).}
	\label{fig7}
\end{figure}

The $s$ dependence in Eq. (\ref{3}) is an experimental finding, in which we have focused only on $T_h(s)$.
Applying Eq. (\ref{9}), obtained in the RM, to the phenomenologically determined Eq. (\ref{4}), we can write
\begin{equation}
\frac{\sqrt{s}^\beta}{T_h(s)} = \frac{1}{T_1} -\frac{n_s}{3}\left( \frac{1}{T_1}-\frac{1}{T_2}\right ) , \label{11}
\end{equation}
where $T_1$  and $T_2$ are defined in Eq. (\ref{5a}). Since $n_s=0$ for proton, we have from Eq.(\ref{9}) $T_q=T_p$, so $T_q(1)=T_1$. Similarly, since $n_s=3$ for $\Omega$, we have $T_s=T_\Omega$, which leads to $T_s(1)=
T_2$.
The LHS of Eq. (\ref{11}) is a representation of the data on the hadrons, while the linear dependence on $n_s$ on the RHS is derived from our model with $T_{1,2}$ referring to quark temperatures. That relationship can be put to test, as we do in Fig. \ref{fig7}, in which Table \ref{tab1} has been used to determine the data points. Evidently, those points are close to the straight line, which is drawn in accordance to the RHS of the above equation. The excellent agreement between data and model is quantified by the ratio (data/model) that shows deviation from 1 to be less than 3\%  for all energies.
 
It is intuitive that the value of $\beta$ is positive, and not large. 
However, it is hard to imagine the origin of how the 66 numbers in Table I can be organized vertically by one number, $\beta$. By comparison, the horizontal organization by two numbers, $T_{1,2}$ is trivial in the RM. All the baryons and antibaryons are produced from a thermal source of $q$ and $s$ quarks and their antiquarks. The RM relates the measured temperatures in the hadronic spectra of $b$ and $\bar b$ to the temperatures of the parton source.

We note that the parton distribution in Eq. (\ref{6}) makes no explicit reference to the quark masses. Although gluons seem to play no explicit role in the recombination formula (\ref{5}), it should be understood that in the formulation of the RM the gluons have always been assumed to convert to quark-antiquark pairs before hadronization. The light and strange quarks exchange energy readily through gluons to form the thermal core.
The {\it a posteriori} knowledge that $T_{q,s}(s)$ are self-similar suggests a common mechanism that generates the thermal core independent of $\sqrt s$, so long as it is high enough. To have that mechanism to start working at  $\sqrt {s_{NN}}$ as low as ~8 GeV was certainly unexpected. Unfortunately, since energy dependence has never been an essential concern of the parton model which is based on the scaling behavior of inclusive distributions, it is questionable whether the mechanism for the production of a self-similar thermal core, characterized by a universal scaling exponent, $\beta$, can be found within the parton model.

\begin{figure}[hpt]
\centering
	\includegraphics[width=0.48\textwidth]{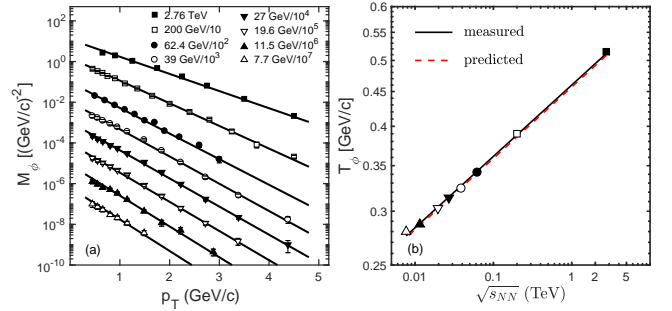}
	\caption{(a) Meson function $M_\phi(s,p_T)$ from data at 0-10\% centrality for all collision energies except 62.4 GeV (0-20\%). Lines are best fits of the data points for $p_T<3$ GeV/c. (b) Points are the inverse slopes $T_\phi(s)$ from the left panel. Solid line is best fit of those measured temperatures, compared to the prediction (dashed) from Eq.(\ref{13}).  }
	\label{fig8}
\end{figure}

Despite the lack of a quantitative model for the production of a thermal core, we want to demonstrate that the study presented here is on the right track --- by a prediction that can actually be verified by existing data. From our knowledge about the $s$ quark source, we can calculate the $p_T$ spectra of the $\phi$ meson. Since $\phi$ consists of $s\bar s$ pair, we modify Eqs. (\ref{6}) and (\ref{7}) for just two partons and get
\begin{equation}
p^0{d{\bar N}_\phi\over dp_T} = A_\phi p_T^2 e^{-p_T/T_\phi}, 	\label{12}
\end{equation}
 where $A_\phi=g_\phi C_s^2$. The  temperature $T_\phi$ is given in the RM by
 \begin{equation}
T_\phi(s) = T_s(s) = T_2 {\sqrt s}^\beta , \quad T_2=0.457\ {\rm GeV/c}.	\label{13}
\end{equation}
 To check this prediction, we define a meson function similar to the baryon function in Eq. (\ref{1})
 \begin{equation}
M_\phi(s, p_T)=\frac{m_T^\phi}{p_T} \left[\frac{dN_\phi}{2\pi p_T dp_T dy}(s, p_T)\right].  \label{14}
\end{equation}
Note that the prefactor above differs from that in Eq. (\ref{1}) because of the difference between Eqs. (\ref{8}) and (\ref{12}). 
We can use the data for $\phi$ distributions in central collisions at various energies \cite{Abelev:2014uua, STAR:2019bjj, STAR:2008bgi} for the quantity inside the square brackets  and obtain Fig.\ref{fig8}(a). The lines are best fits for the region $p_T<3$ GeV/c; yet by extension they show good fit of all the experimental points, exhibiting exponential behavior  up to the maximum  $p_T$ measured. We therefore can use the formula
\begin{equation}
M_\phi(s,p_T) = A_\phi(s) \exp[-p_T/T_\phi(s)]    \label{15}
\end{equation}
to describe that behavior and show the {\it measured} values of $T_\phi(s)$ as points in Fig.\ref{fig8}(b). The solid line is a best fit of all those points that yields
\begin{equation}
T_\phi(s) = T_\phi(1){\sqrt s}^\beta, \quad  T_\phi(1)=0.46\ {\rm GeV/c}, \quad  \beta=0.105.  \label{16}
\end{equation}
It agrees excellently with the dashed line that represents our prediction according to Eq.(\ref{13}). We have hereby presented evidence in support of the reliability of  the power-law behavior that we have found to describe the thermal source of light and strange quarks.

\section{summary}
\label{sec5}
In conclusion, we have found in heavy-ion collisions a quantity that can reasonably be called {\it measured} temperature, which is the inverse slope of an exponential behavior in transverse momentum, independent of any theoretical model. That temperature $T_h(s)$ possesses a scaling behavior, ${\sqrt s}^\beta$, that is universal across all hadron types $h$. The observed universality is valid only for central collisions with violation beginning to occur at 30-40\% centrality.  The scaling exponent $\beta$ is indeed a constant number, $0.105$. Finding a universal constant in high-energy collisions rarely occurs. We do not have at this point any explanation for its existence. No model has been used in our data analysis. We present this finding as a remarkable phenomenological property in heavy-ion collisions to be considered by the community for further experimental and theoretical investigation.

We end by making some comments looking forward. They are model dependent and should therefore not affect the credibility of the phenomenological features presented above, but may be regarded as our speculation on what those features may imply from the present viewpoint. The central element of the results of our study is that hyperon production share the same properties as proton production; thus variations in strangeness and large differences in baryonic masses are not distinctive barriers for universality. Furthermore, baryons and anti-baryons share essentially the same characteristics. These are not the properties usually ascribed to an equilibrated hadronic gas assumed to form at the end of hydrodynamical expansion following heavy-ion collisions. However, for a partonic system consisting of light ($q$) and strange ($s$) quarks and their antiquarks ($\bar q$ and $\bar s$), those properties are consequences of partonic interactions through gluon exchanges, $q\bar q$ and $s\bar s$ annihilation and creation, provided that the system is in a state of thermal energy higher than the rest masses of quark pairs. From Table 1 we see that all values of $T_h$ are well above the $q\bar q$ and $s\bar s$ masses ($m_q=5$ MeV, $m_s=100$ MeV) even at the lowest BES energy. Thus we are led to interpret the observed universality behavior as being due to the formation of a hot thermal system of partons, from which baryons and anti-baryons are formed through recombination  \cite{Hwa:2018qss, Das:1977cp, rm1, rm2, Molnar:2003ff, Fries:2003vb, Greco:2003xt}.  

\section*{Acknowledgements}
This work was supported in part by the National Natural Science Foundation of China under Grants Nos. 11905120, 11947416 and the Natural Science Foundation of Sichuan Province under Grant No. 2023NSFSC1322. We thank Dr. Wenbin Zhao for helpful discussion.








\end{document}